\documentclass[]{aastex631}

\received{February 19, 2024}
\revised{April 4, 2024}
\accepted{April 11, 2024}

\begin{document}


\title{Jupiter Co-Orbital Comet P/2023 V6 (PANSTARRS):\\
Orbital History and Modern Activity State}

\correspondingauthor{Theodore Kareta}
\email{tkareta@lowell.edu}

\author[0000-0003-1008-7499]{Theodore Kareta}
\affiliation{Lowell Observatory\\
1400 W. Mars Hill Road \\
Flagstaff, AZ, USA, 86001}

\author[0000-0003-2152-6987]{John W. Noonan}
\affiliation{Physics Department\
Edmund C. Leach Science Center\\
Auburn University\\
Auburn, AL 36849, USA}

\author[0000-0001-8736-236X]{Kathryn Volk}
\affiliation{Planetary Science Institute \\
Tucson, AZ, USA}

\author[0000-0001-6350-807X]{Ryder H. Strauss}
\affiliation{Department of Astronomy and Planetary Science\\ Northern Arizona University\\ PO Box 6010, Flagstaff, AZ 86011, USA}

\author[0000-0003-4580-3790]{David Trilling}
\affiliation{Department of Astronomy and Planetary Science\\ Northern Arizona University\\ PO Box 6010, Flagstaff, AZ 86011, USA}

\begin{abstract}
The discovery of the transient Jupiter co-orbital comet P/2019 LD2 (ATLAS) drew significant interest. Not only will LD2 transition between being a Centaur and a Jupiter Family Comet (JFC) in 2063, the first time this process can be observed as it happens, it is also very active for its large heliocentric distance. We present observations and orbital integrations of the newly discovered transient Jupiter co-orbital comet P/2023 V6 (PANSTARRS), the second such object known. Despite similar modern orbits, V6 is significantly (15$\times$) less active than LD2 and most JFCs as determined via $Af\rho$ measurements at the same $R_H$. We find V6 is co-orbital between 2020 and 2044, twice the duration of LD2, but it will not become a JFC soon. We interpret these differences in activity as evolutionary, with V6 having lost a significant fraction of its near-surface ice compared to LD2 by previously being warmer. While V6's pre-encounter orbit was somewhat warmer than LD2's, future thermal modeling will be needed to understand if this can explain their differences or if a more significant difference further into the past is required. This is more evidence that LD2 is a pristine and ice-rich object, and thus it may display very strong activity when it becomes a JFC. We use the differences between V6 and LD2 to discuss the interpretation of cometary activity at large heliocentric distances as well as the small end of the cratering record of the Galilean Satellites. Continuing observations of both objects are highly encouraged.
\end{abstract}


\section{Introduction} \label{sec:intro}
Centaurs, Solar System small bodies on chaotic orbits between Jupiter and Neptune, are a transient population fed by the inward orbital migration of Kuiper Belt objects that in turn feeds the short period Jupiter Family comets (JFCs) (see, e.g., \citealt{DiSisto:2020,Fraser:2022}). Although the fraction of their total lifetime spent as JFCs is remarkably small, it is this short period when nearly all observations of the typical active small object are obtained; the decreased distance to the Sun facilitates higher surface temperatures and thus active sublimation of embedded water ice and the production of large comae. However, cometary activity does not start when an object enters the Jupiter family -- a significant heliocentric-distance-dependent fraction of the Centaurs are active \citep[e.g.][]{2009AJ....137.4296J, 2019A&A...621A.102C, 2021PSJ.....2..155L}, showing persistent or transient mass loss well beyond $10$ AU. The nature of this distant cometary activity is an area of much active research. Some Centaurs display stable activity like `typical' comets, whereby their bulk brightness and dust production rates are correlated with heliocentric distance \citep{2021PSJ.....2...48K}, while others display stochastic outburst-dominated modes of activity \citep{2010MNRAS.409.1682T} that has not been observed in other populations. It is thus challenging to assess how this phase of activity that occurs on many Centaurs alters interpretations of the activity states of the JFCs. Current estimates (see, e.g., \citealt{2017ApJ...845...27N}) suggest that the JFCs can survive $300-800$ orbits with perihelia inside the water sublimation region ($q < 2.5$ AU) before they go dormant or disintegate, but the importance of their time with larger perihelia is challenging to constrain with current population models. Do some objects move through the Centaur region significantly faster than others such that they reach the JFCs with more of their original volatile content to sublimate, or are all objects affected by their time among the Giant Planets at a roughly similar level? Furthermore, would it be possible to separate any differences in volatile contents from other effects like size, rotation state, or obliquity?

The best studied (and most active) active Centaurs like 29P/Schwassmann-Wachmann 1 or 174P/Echeclus are almost all much larger ($D>30$ km, \citealt{2021PSJ.....2..126S,2024MNRAS.527.3624P}) than any known JFC , thus making a direct comparison to the JFC population challenging. Advances in the sensitivity of asteroid and comet surveys have begun to find fainter -- and often smaller -- active Centaurs, but sample sizes remain small and direct detections of nuclei are increasingly challenging at larger distances. Two active Centaurs within the size range of most JFCs are P/2019 LD2 (ATLAS, hereafter just ``LD2"; \citealt{2021PSJ.....2...48K, 2021A&A...650A..79L, 2021AJ....161..116B}) and 39P/Oterma \citep{2023PSJ.....4..208P}. LD2 has a nuclear radius  $r_{nuc} < 1.2$ km based on non-detections and assuming a visual albedo of $5\%$ \citep{2021PSJ.....2...48K} and Oterma has a nuclear size between $2.2-2.5$ km based on groundbased observations obtained while minimally or inactive \citep{2023PSJ.....4..208P}. For context, surveys of comet sizes (see, e.g., \citealt{2017AJ....154...53B} find an average size for the JFCs of $r\sim0.65$ km with smaller sizes being more common than larger ones. It might be expected that a Centaur selected at random from the whole population might be slightly larger than a similarly randomly selected JFC as the JFC would have spent more time in a warmer environment, encouraging more ice sublimation and thus erosion of the surface -- though the exact magnitude of this effect is of some debate. Observational biases account for some of larger sizes of typical observed Centaurs, as the Centaurs are further away and thus the same limiting magnitudes would detect larger targets.

That said, some unlucky objects have spent time as JFCs only to be flung back out into a Centaur orbit -- Oterma is the classic example of this, as it was a Centaur that was tossed inwards in 1937, was discovered as a JFC in 1943, and then tossed back outwards in in 1963 \citep[see, e.g.,][]{Fernandez:2018}. The object's minimal activity in its current Centaur orbit is primarily attributed to this passage through the Inner Solar System (see a longer discussion of this in \citealt{2023PSJ.....4..208P}). Comets with more distant perihelia appear to have higher $CO/CO_2$ ratios than those closer in \citep{2022PSJ.....3..247H}, and Oterma's $CO$ production rate is lower than expected \citep{2023PSJ.....4..208P} in a way that is attributable to its dynamical history. Clearly for some objects, it is possible to better contextualize their modern activity through an understanding of the object's past thermal evolution. It is thus critical to combine observations with an assessment of an object's orbital history in order to put the object in it's proper context.

Unlike Oterma, LD2 has likely not spent appreciable time in the inner Solar System yet \citep{2020ApJ...904L..20S}. The object is currently co-orbital with Jupiter, but only temporarily between close encounters with the gas giant in 2017 and 2029 \citep{2020RNAAS...4...74K, 2020ApJ...904L..20S, 2021Icar..35414019H}. LD2's high activity in this current orbit, lack of signs of activity in precovery images \citep{2021PSJ.....2...48K}, and past orbital trajectories \citep{2020ApJ...904L..20S} all point towards LD2 maintaining much of its original volatile content -- at least as much as an object at these heliocentric distances can maintain. While activity that strengthened towards perihelion suggests comet-like sublimation as the primary activity driver, the initiation mechanism(s) for activity on small bodies in this region remain elusive to capture and difficult to characterize.

Simultaneous investigation of the orbital dynamics of the evolution of the broader Centaur population has also shown that there is a ``gateway", a set of orbital properties that facilitates the transition of Centaurs to JFCs in dynamic evolution models \citep{2019ApJ...883L..25S}. Activity within and near this gateway, like that seen on 29P and LD2, indicates that substantial thermal and geological evolution (i.e., sublimation) can take place on these object prior to entering their more frequently studied JFC orbits \citep{2023ApJ...942...92G}, and in turn may affect their observed composition, structure, and skew our interpretation of their perceived bulk compositions inherited from the natal solar system. As we make progress in understanding the modes in which Centaurs are active, it is thus critical to compare apples to apples: given all of these challenges and complicating factors, comparisons of objects in more similar orbits are more likely to be meaningful than those in completely different ones.

In this paper, we present orbital integrations and visible-wavelength imaging of a newly discovered active outer Solar System object, the Jupiter co-orbital comet P/2023 V6 (PANSTARRS), hereafter referred to as ``V6". The primary aim of this paper is to diagnose the object's activity state, contextualize it within its short- and long-term orbital evolution, and use this to compare to other low-perihelion Centaurs and high-perihelion JFCs -- including and primarily P/2019 LD2 (ATLAS). In Section \ref{sec:obs}, we detail the observations we obtained to both provide astrometry to improve the object's orbit and to assess its modern activity state. In Section \ref{sec:dyn}, we discuss the path that took V6 into its current orbital state and begin to comment on its likely thermal history. In Section \ref{sec:disc}, we synthesize observations and dynamics to compare V6 to other comets at the same distance -- including LD2.
 
\section{Observations of P/2023 V6} \label{sec:obs}
\subsection{Observations Description and Spatial Profiles}
Observations of P/2023 V6 were obtained with the Lowell Discovery Telescope's (LDT) Large Monolithic Imager (hereafter `LMI', \citealt{2014SPIE.9147E..2NB}) on 22 November 2023 UTC to both obtain astrometry to refine the objects orbit (to facilitate the dynamical analyses described in Section \ref{sec:dyn}) and to understand its current activity state and compare it to objects in similar orbits and thermal environments. A series of forty 120-second exposures were obtained through the $VR$ filter (similar to other $g+r+i$ high throughput filters) interspersed between other targets throughout the second half of the night in $2\times2$ binning (a pixel scale of $0.24\arcsec$). The $VR$ filter, despite its broader bandpass, is sufficiently close in central wavelength to the Sloan $r$ filter such that it can be calibrated against the PANSTARRS $r$ magnitudes of field stars with only a few percent loss in precision. Photometric calibration as well as astrometric registration of images was accomplished using PhotometryPipeline \citep{2017A&C....18...47M}. We only utilized stars with visible colors similar to those of the Sun in our photometric calibration to ensure our calibrators were similar in spectral behavior to our target and thus that our $VR$-derived magnitudes were calibrated as accurately as possible. This calibration method and the high throughput of the $VR$ filter has made it an effective tool for characterizing relatively faint Solar System targets (see, e.g., \citealt{2020RNAAS...4..101Y}'s recovery of 12P/Pons-Brooks) at the LDT.

\begin{figure}[ht!]
\plotone{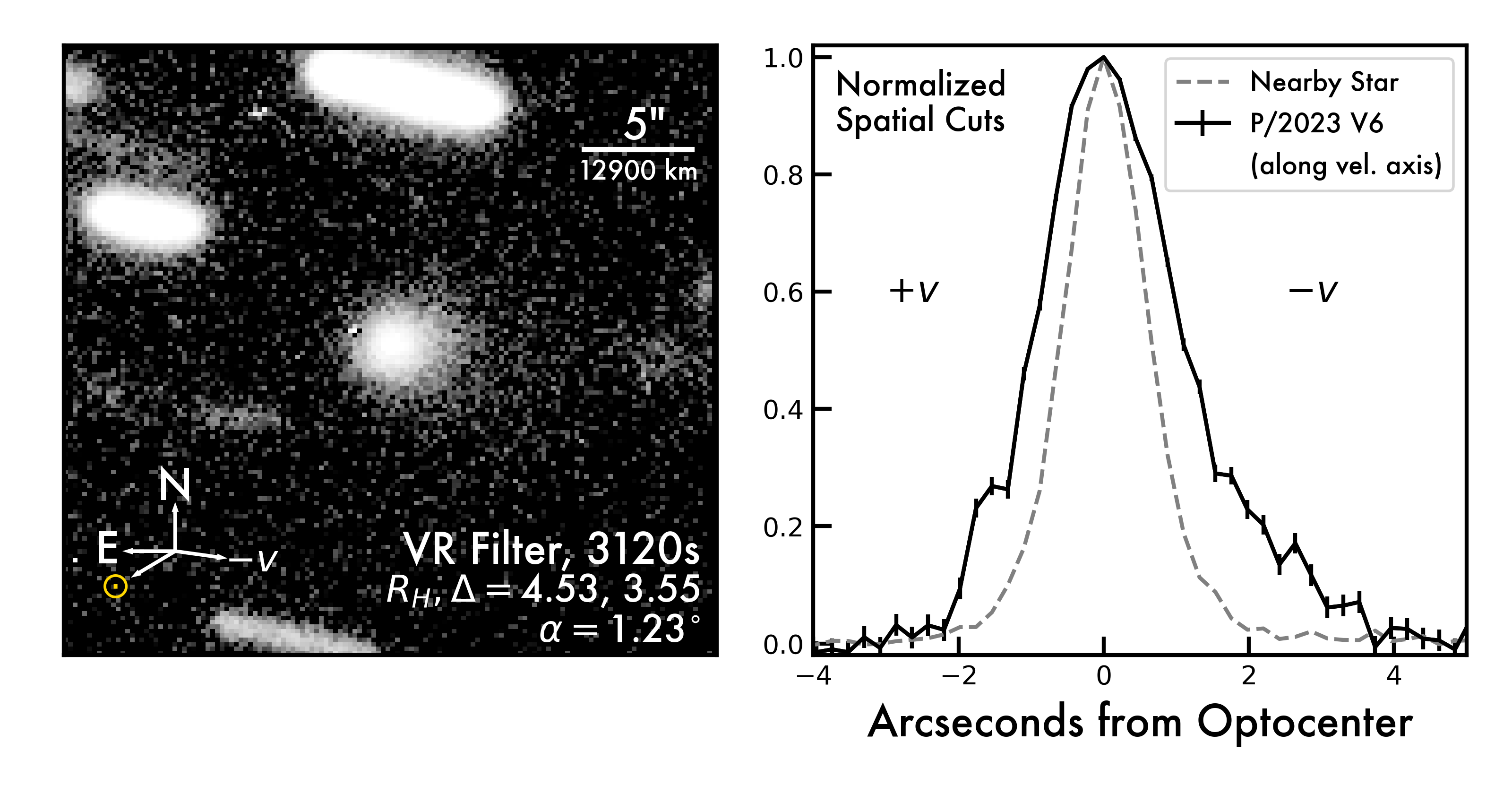}
\caption{Left: A stack of a subset of our Lowell Discovery Telescope/LMI imaging of P/2023 V6 (PANSTARRS) on 22 November 2023 totaling 3120s (or $\sim0.86$ hours) shown in a logarithmic stretch between the peak brightness of the comet and 2$\%$ of it. A legend in the lower left shows the direction towards the Sun (indicated with a yellow circle-dot symbol) and the direction opposite the objects velocity (denoted with a $-v$). Right: spatial cuts along the comet's velocity axis of the left panel image are shown, highlighting the comet's weak but clear extension in its anti-velocity direction. The low phase angle and moderate heliocentric distance of these observations means that most of the tail is radially away from the observer. As mentioned in the text, the leading profile of the spatial cut can be fit with a power-law of slope $-0.9\pm0.2$, consistent with the expectation for a comet with relatively stable activity.
}
\label{fig:images}
\end{figure}

A stack of the subset of our imaging data in which V6 was at least several arcseconds away from any nearby faint star (26 of the 40 total frames) is shown in Figure \ref{fig:images} alongside a series of spatial ``line" cuts along the direction of V6's on-sky movement. At the time of our observations, V6 was a relatively condensed target approximately $\sim4\arcsec$ across but still clearly extended compared to nearby stars in the $\sim1.1-1.2\arcsec$ seeing. V6 was clearly more extended in the anti-velocity direction (relatively similar to the anti-Sun direction), almost certainly driven by the existence of a dust tail, but its tail is not extremely prominent. However, due to the considerable heliocentric distance ($R_H=4.53$ AU) and low phase angle ($\alpha=1.23^{\circ}$) at which V6 was observed, the direction of the dust tail was likely almost entirely radially outwards -- and thus much of the brightness of the tail is `behind' the central coma, contributing to its brightness. A power-law fit to the spatial profile's sunward edge results in a slope of $-0.9\pm0.2$, consistent with the expectation for an idealized cometary coma in steady state mass loss \citep{1984AJ.....89..579A}, though we note that future deeper imaging with better seeing than ours might be able to better assess the steadiness of its activity. That said, the magnitudes reported to the Minor Planet Center between mid-October and mid-November 2023 are relatively stable near $m_V \sim21$, so there is no evidence of large outbursts or changes in activity that might argue against steady-state activity.

\subsection{Brightness and Activity State}
Based on the spatial extent of the object seen in our deep stack and spatial cut analyses, we extracted aperture photometry of the target with a 9-pixel radius, equivalent to $2.16\arcsec$ or $\sim5600$ km. Including only frames where V6 was not very close to a nearby field star, we obtain a mean brightness of $m_{VR/r}(r_{ap.}=2.16\arcsec)=20.90\pm0.03$. No significant variations were seen in the brightness of V6 over the $\sim5$ hours between our first and last observation. We can convert this brightness to an $Af\rho$ value \citep{1984AJ.....89..579A}, a measure of cometary activity strength with units of length which (in theory) is only minimally affected by aperture size for comets with steady activity, retrieving $Af\rho = (32\pm1)$ cm after correcting to a phase angle of $0^{\circ}$ \citep{1998Icar..132..397S}. Observations of P/2019 LD2 (ATLAS) at a very similar heliocentric distance ($R_H = 4.58$ AU) in July 2020 reported in \citet{2021PSJ.....2...48K} found $Af\rho=(484\pm20)$ cm for that object, about fifteen times higher than V6. Observations of both objects were obtained a few months after their respective perihelia, another indication that their bulk thermal states should be similar. Those authors chose their aperture size based on where the power-law slope of LD2's coma was similar to the idealized $-1$ value, a constraint which is satisfied for our $r_{ap}=2.16\arcsec$ extraction as noted above. The observing geometry, as described above, means that at least some of V6's brightness is due to the tail, and thus our $Af\rho$ measurements are likely a slight over-estimate. While there are many factors which affect an object's $Af\rho$ value, V6 is losing much less mass per second than LD2, despite their very similar modern orbits. This could be due to a difference in overall activity strength, V6 being significantly smaller than LD2, or a combination of these factors; we return to this comparison in Section \ref{sec:disc}.

We also made observations of V6 on two subsequent nights at the LDT, 20 and 29 December, when seeing was worse ($\sim2.0$ arcseconds FWHM) and thus not suitable to morphological analyses. Assuming that a similar extraction aperture as before was still appropriate, a stack of eight $300$s images on Dec. 20 resulted in $m_{VR/r}(r_{ap.}=2.16\arcsec)=21.44\pm0.04$ and a stack of five $300s$ images on Dec. 29 resulted in $m_{VR/r}(r_{ap.}=2.16\arcsec)=21.56\pm0.06$. These values are somewhat dimmer compared to the Nov. 20 observations, but the differences in phase angle ($1.23^{\circ}$ vs. $5.82^{\circ}$ vs. $7.62^{\circ}$, chronologically) account for most of the variation if a standard $0.04$ mags/degree phase curve is assumed. These magnitudes convert to $Af\rho$ values of $25\pm1$ cm and $24\pm1$ cm on Dec. 20 and 29 respectively after a phase angle correction to $\alpha=0^{\circ}$. While these values are slightly smaller than our Nov. 20 observation, and thus may be interpreted as the object's activity having weakened, we remind the reader that the worse conditions on these two dates as well as decreased contamination by the tail both likely contribute at some level. Therefore, and combined with the stability of brightness reports to the MPC as mentioned above, we conclude that V6's activity throughout 2023 November through December was relatively stable or decreased slightly as the object receded from perihelion. Stable and continuing activity is another property that V6 shares with LD2, adding credence to the comparison in the previous paragraph and heightening the discrepancy in their relative activity strengths. We compile our $Af\rho$ values for V6 along with those of LD2 \citep{2021PSJ.....2...48K} in Table \ref{tab:vals}, which also includes $Af\rho$ values for several other comets at similar heliocentric distances described in more detail in Section \ref{sec:disc}.

\section{Past and Future Orbital Evolution} \label{sec:dyn}
\subsection{Recent Changes and Comparisons to LD2}
The current orbit of P/2023 V6 (PANSTARRS), specifically its proximity and period similarity to Jupiter, suggests that substantial orbital evolution on relatively short timescales is to be expected. As a result, to properly categorize the object and to contextualize observations of its current activity state, orbital integrations are required. To assess the potential orbital evolutionary pathways of P/2023 V6, we performed simulations using the \textsc{rebound} orbital integration package \citep{2012A&A...537A.128R} accessed through Python with the IAS15 integrator \citep{2015MNRAS.446.1424R}. IAS15 is an adaptive-timestep integrator, which is critical for being able to properly integrate through close planetary approaches as is expected to be common for V6. Given the significant uncertainties in V6's orbital elements, we used a multivariate sampling scheme appropriate for use with covariance matrix of the JPL orbit fit as queried from the NASA JPL Small Bodies Database. This general approach thus accounts for challenges with integrating through close encounters, the non-zero current uncertainties in this object's orbit, and how close encounters can cause that uncertainty to grow non-linearly with time. As a result, this kind of procedure has been used successfully to the study the dynamics of notable Centaurs in several previous works (see, e.g., \citealt{2019AJ....158..255K,2019ApJ...883L..25S, 2020ApJ...904L..20S, 2021Icar..35414019H}). Handling uncertainties properly is even critical to interpreting the output of individual integration outputs as the onset of chaos \citep{2020MNRAS.497L..46M}, and thus when integrations begin to some of their physicality, can only be assessed through the divergence of the clones from each other. We conducted two sets of integrations: the first included 1000 orbital `clones' that were integrated forwards and backwards 70 years whose positions were written out every $\frac{1}{50}$ of a year, and the second included 100 clones integrated forwards and backwards 500 years whose positions were written out every $\frac{1}{10}$ of a year. These integrations were designed to both probe the objects modern co-orbital state and the object's longer-term bulk evolution.

\begin{figure}[ht!]
\plotone{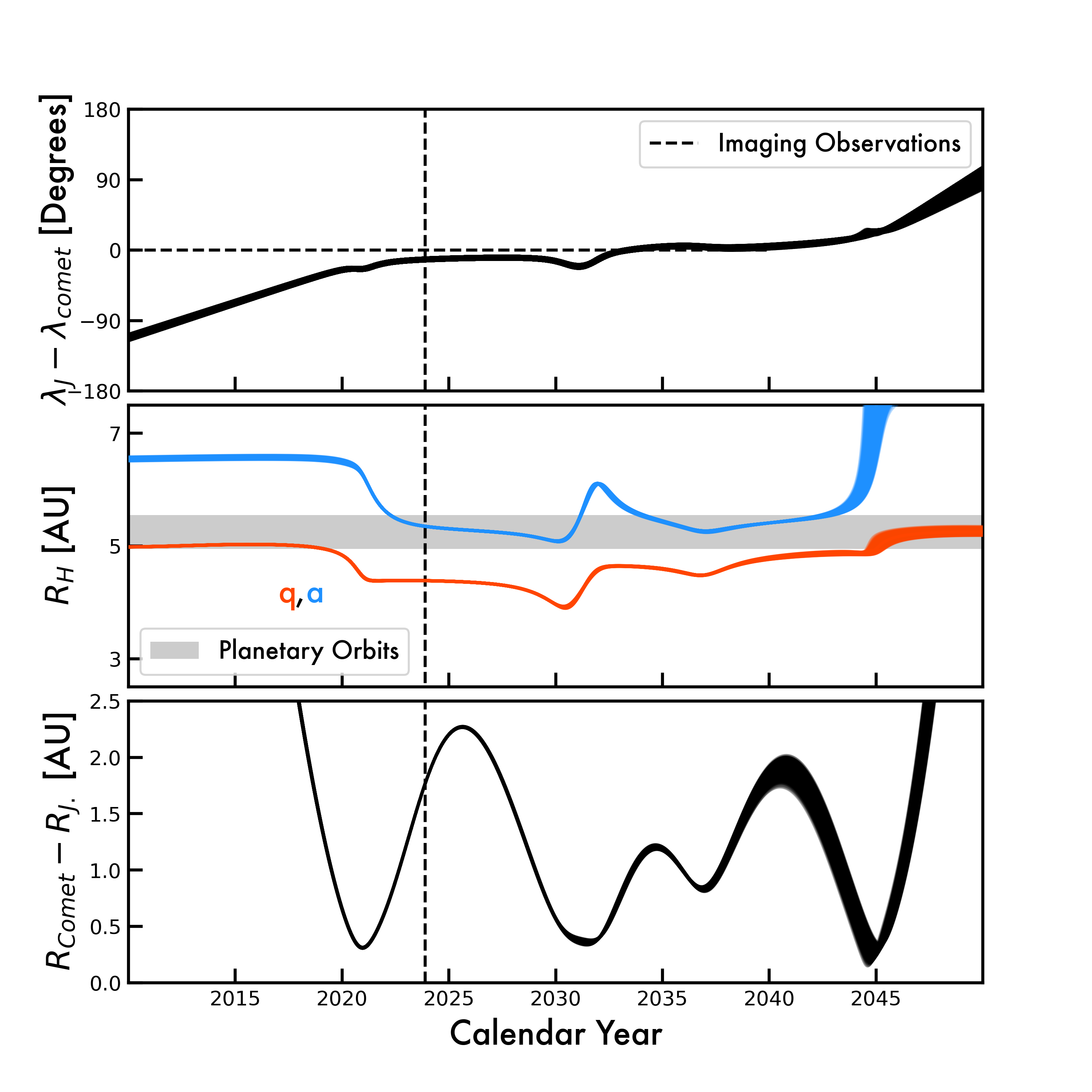}
\caption{The orbital evolution of P/2023 V6 (PANSTARRS) between the years 2010 and 2050 is shown, along with that of 1000 orbital `clones' consistent with the object's orbit as described in the text. Top: the evolution of the difference in orbital longitude of V6 and Jupiter is shown in black. Stable values of this angular difference (e.g. where the curves are approximately `flat') are when the two objects have very similar orbital periods, and are thus co-orbital. Middle: the evolution of V6's semi-major axis ($a$) in blue and perihelion ($q$) in orange are shown. Bottom: The distance between the comet and Jupiter is plotted in black. The sudden jumps in $q$ and $a$, like those in 2020, 2032, and 2044, are from close-encounters with Jupiter. The 2044 encounter is by far the deepest, and where the common path of the clones diverge due to cumulative uncertainties in the object's orbit becoming substantial, and is discussed at length in the text.
}
\label{fig:researchnote}
\end{figure}

Figure \ref{fig:researchnote} shows the very recent and near-future orbital evolution of V6 between the years 2010 and 2050 (the short-duration 1000 clone integration described in the previous paragraph). V6 is shown to be co-orbital with Jupiter from the end of 2020 onwards, with most clones remaining co-orbital until a close encounter with Jupiter in 2044, described in depth below. V6's entrance into its current orbit might is somewhat similar to how LD2 entered its very similar modern orbit, but there are key differences worth discussing. First, V6 stays co-orbital with Jupiter at least twice as long as LD2 is expected to. While LD2 stays co-orbital for a single jovian orbit ($\sim12$ years), the vast majority of clones of V6 last for two jovian orbits or more. V6 still has another close encounter with Jupiter one orbital period after it entered its co-orbital state, but this encounter modifies V6's orbit only somewhat -- it does not eject it out of a co-orbital state but instead moves it into a lower-eccentricity co-orbital state. Second, their pre-encounter orbits were also different. While V6's perihelion prior to orbital insertion was very nearly $\sim5.0$ AU exactly, on the interior edge of Jupiter's heliocentric distance range, LD2's perihelion was at the outermost edge of Jupiter's range at $q\sim5.5$ AU. The close encounter that pushed V6 into its current orbit thus happened at higher orbital velocities than what happened to LD2 as \textit{both} bodies were at or near perihelion when the encounter occurred.

\begin{figure}[ht!]
\plotone{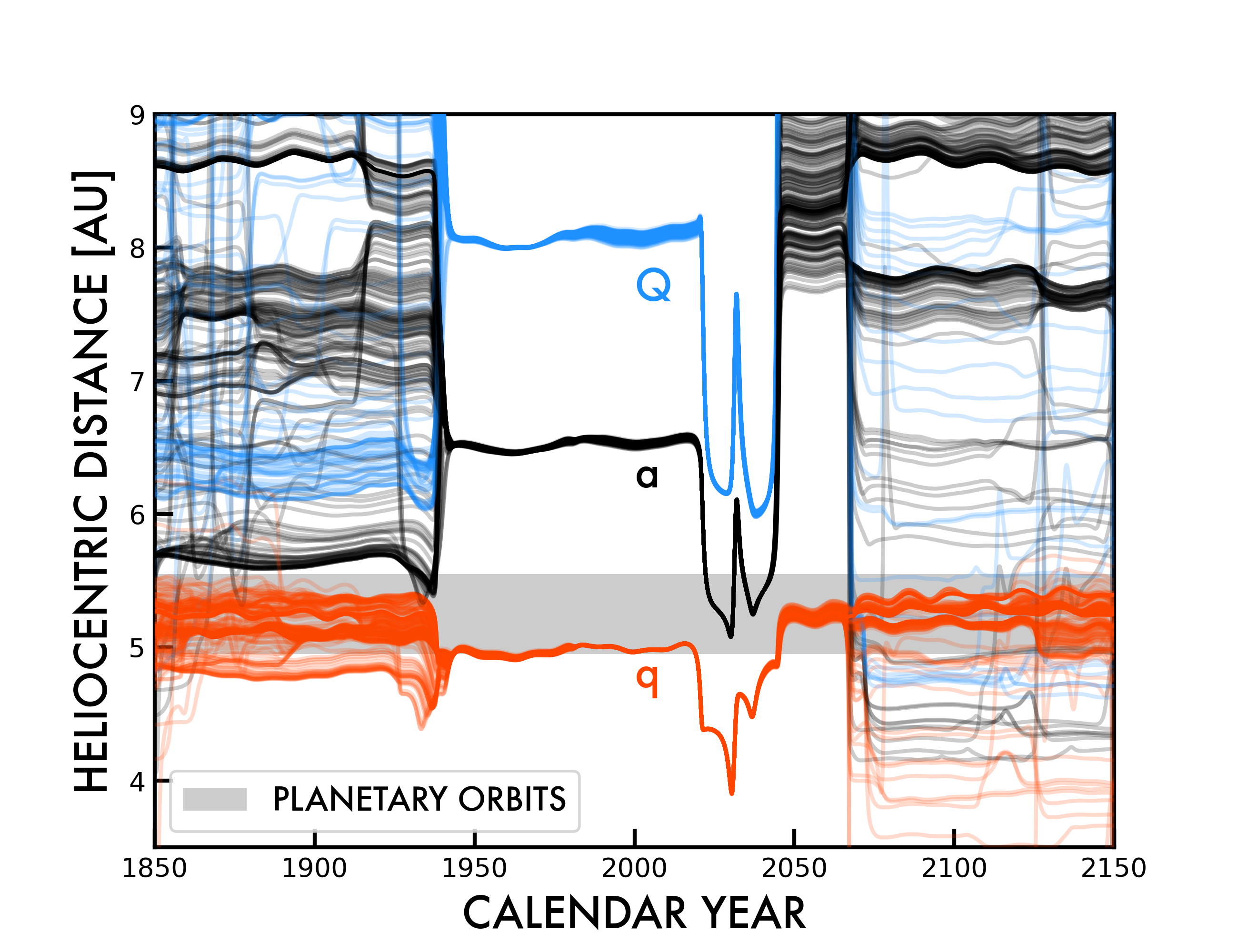}
\caption{A long-term integration of P/2023 V6 (PANSTARRS) and one hundred orbital clones is shown, essentially a `zoom out' of the bottom panel of Figure \ref{fig:researchnote}, with V6's perihelion, semimajor axis, and aphelion plotted in orange, black, and blue respectively. The object's 2022 to 2044 co-orbital status is book-ended by long periods where the object has a perihelion in the vicinity of Jupiter's orbit, but with a significantly larger semimajor axis than its current value. The orbital integrations begin to decohere prior to a close encounter with Jupiter in $\sim$1942 and after a close encounter with Jupiter in $\sim$2067.
}
\label{fig:longterm}
\end{figure}

Figure \ref{fig:longterm} shows the evolution of V6's perihelion, semimajor axis, and aphelion distances between the years 1850 and 2150 (the long-duration 100 clone integration). V6's co-orbital phase is, as expected, transient. Prior to the 2020 close encounter which pushed the object into its current configuration, V6 was in a relatively stable orbit with $q\approx5$ AU and $a\approx6.5$ AU since at least the early 1900s. This value for V6's perihelion is too low to be considered a Centaur in some definitions, in which case `high perihelion JFC' might be preferred. No signs of rarer dynamical phenomena, such as a phase as a transient moon or Trojan, were noted in the integrations. These longer-duration stays around Jupiter are probably easier to initiate for objects which encounter Jupiter at aphelion \citep{1995Icar..118..155B}, and thus lower relative velocities. This is not the case for V6 (or LD2) which encounter Jupiter with higher relative velocities near perihelion. 

The pre-encounter orbit for LD2 was slightly larger -- a perihelion just exterior to Jupiter ($q\approx5.5$ AU) and a semimajor axis about an AU further out ($a\approx7.6$ AU) (see, e.g., \citealt{2020ApJ...904L..20S}). \footnote{Despite their similar orbital evolutions, LD2's pre-encounter orbit would thus be categorized as a Centaur. This highlights the challenges in classifying objects near the edges of dynamical boundaries, especially those which are moving back and forth between groups.} LD2's pre-encounter orbit was also preceded by some hundreds of years of stability despite the possibility of close encounters with Jupiter. Recent work \citep{2024ApJ...960L...8L} has shown that decreases in a Centaur's semimajor axis ($a$) -- and thus increases in average surface temperature -- can initiate mass loss at that object. Both LD2 and V6 thus experienced this drop in $a$, and thus would be expected to be active. However, that same work \citep{2024ApJ...960L...8L} showed that there is typically a lag between the orbital change and the onset in activity averaging a hundred years with significant variation. LD2 had a larger semimajor axis before 1850 \citep{2020ApJ...904L..20S, 2024ApJ...960L...8L}, so activity might have already been ongoing at it prior to its 2017 close encounter, but no positive detections of the object have been found prior to the encounter \citep{2021PSJ.....2...48K}. V6's current orbital fit combined with its recent encounter history does not facilitate accurate integrations as far back in time as LD2's, so the previous jumps in $a$ that V6 may have encountered -- and thus any sense of whether it may have been active at these larger distances -- is unclear at present. Considering the object's current faintness, it seems unlikely that precovery data will be plentiful.

As for the thermal states of these two objects, V6 has been slightly warmer for at least a century or so -- assuming similar albedos, V6 had surface temperatures at perihelion about $\sim10\%$ higher than LD2 at most. Granted, surface temperatures at these distances even for low-albedo objects are not high -- the difference should be about $\sim10$K or less. Would this slight difference in surface temperatures for a century or so even be enough to create a significant difference in their near-surface and interior thermal states? If the $\sim100$ year delay between orbital change and onset of activity mentioned above can be taken as a typical amount of time for one of these objects to fully respond to a change in thermal input, V6 and LD2 have been different for long enough for some differences in their interiors to be expected. If some of the variation in lag times seen in \citet{2024ApJ...960L...8L} was due to variation in the sizes of the Centaurs considered (one might imagine that a $r\sim30$km Centaur responds to changes in heating slower than a $r\sim1$ km Centaur), one might expect an event shorter lag for V6 and LD2 due to their smaller sizes than the typical known Centaurs. A complicating factor worth discussing, however, is that Solar insolation can only have affected the upper layers of the comet within a few thermal skin depths, or just a few to ten meters \citep{2015Sci...347a0709G}. Even for the JFC 67P/Churyumov-Gerasimenko, in a significantly warmer orbit, erosion rates are only about a skin depth ($\sim1$m) per orbit \citep{2015A&A...583A..34K} The differences in activity strengths in these objects, assuming that they are in someway related to their different orbital histories, might result from differences in just the outermost parts of their nuclei as opposed to differences in their deep interiors. A future study (once V6's orbit is further refined and its surface albedo can be better estimated) to model and compare the two object's thermal states and thus better contextualize their activity strengths could be useful in assessing this. The chaotic nature of the evolution of comet and Centaur orbits means that one can only ever constrain the recent differences between objects, and that their past orbital and thermal evolution -- beyond where their current orbital solutions decohere -- might have created even more variation in their internal thermal states and volatile abundances in their outer layers. We discuss this further in Section \ref{sec:disc}.

\subsection{Investigating V6's 2044 Encounter with Jupiter}
The proximity to Jupiter and the relatively large uncertainties in the current orbital parameters act in tandem to throw the test clones into chaos relatively quickly; as such we only reverse integrate the particles 200 years into the past and forward integrate them 200 years into the future before they become chaotic. As can be seen in Figure \ref{fig:researchnote} the clones follow a tightly constrained Jupiter co-orbit until 2044, at which point a deep Jupiter encounter drastically alters their semimajor axis and perihelion. This 2044 Jupiter encounter was investigated further at a smaller timestep using the simulation data just prior to the encounter as the initial conditions, with the Galilean moons (Io, Europa, Ganymede, and Callisto) added to the simulation. With the current set of clones populated from the (significant) orbital uncertainties there is no indication that a Jupiter or Galilean collision is likely. After approximately 30 years in a high perihelion JFC orbit ($a$ = 9.0, $q$=5.3), V6 will have another Jupiter encounter in late 2067, which is on average less deep than the 2044 encounter. By this point in the simulation the clones have already begun to spread out in parameter space, so the exact times of the encounter vary.

In the next 50 years following this second encounter one evolutionary path is statistically dominant. In this path, 83.7\% of the P/2023 V6 clones are injected into orbits possessing a median and mean $a$ of 9.27 and 9.15 au, respectively. The similarity of the mean and median suggests a normal distribution with few outliers. These clones have perihelia between 5.26 and 5.43 au, with a group perihelion median/mean of 5.30/5.32 au. The remaining 16.3\% of clones are split between having median semi-major axes smaller than 8.0 au or larger than 10 au (15.9\% and 0.4\%). The inner group that ends up inside of the 8.0 au barrier show median/mean $a$ values of 7.21/7.03 au, while the remaining particles outside of 10 au is (median/mean 10.20/10.54 au) are due to four clones that scattered outward following the encounter. In summary and unlike LD2, there is no indication that V6 will enter a JFC-like orbit within the next century and will instead remain in similar or somewhat more distant orbits.

\section{Discussion} \label{sec:disc}
The primary aim of our study into P/2023 V6 (PANSTARRS) is to understand how its modern activity and orbital evolution are related and to compare with objects in similar orbits. While the typical dynamical lifetimes for JFCs are of order $10^5$ years, their active lifetimes are likely a few thousands to tens of thousands of years (see, e.g., \citealt{1994Icar..108...18L}). Thus any individual area of orbital phase space must be populated with objects of different \textit{physical} ages -- how long and how far they have been inside of each of the respective ice lines, and thus how much of their volatile content has been lost -- and thus very different activity strengths. V6's pre-encounter Centaur-like orbit does not necessarily mean that at some point in the more distant past, where orbital integrations cannot determine the object's individual history absolutely, it was not in a substantially warmer environment (see again \citealt{2023PSJ.....4..208P}). Low eccentricity orbits near Jupiter can move objects inwards and outwards (\citealt{2019ApJ...883L..25S}, see also \citealt{1995Icar..118..155B}), so a primary aim of this Discussion is to try to put the pieces together to tell a complete story of how V6 is acting and where it has been.

\subsection{Activity Strength}
\begin{figure}[ht!]
\plotone{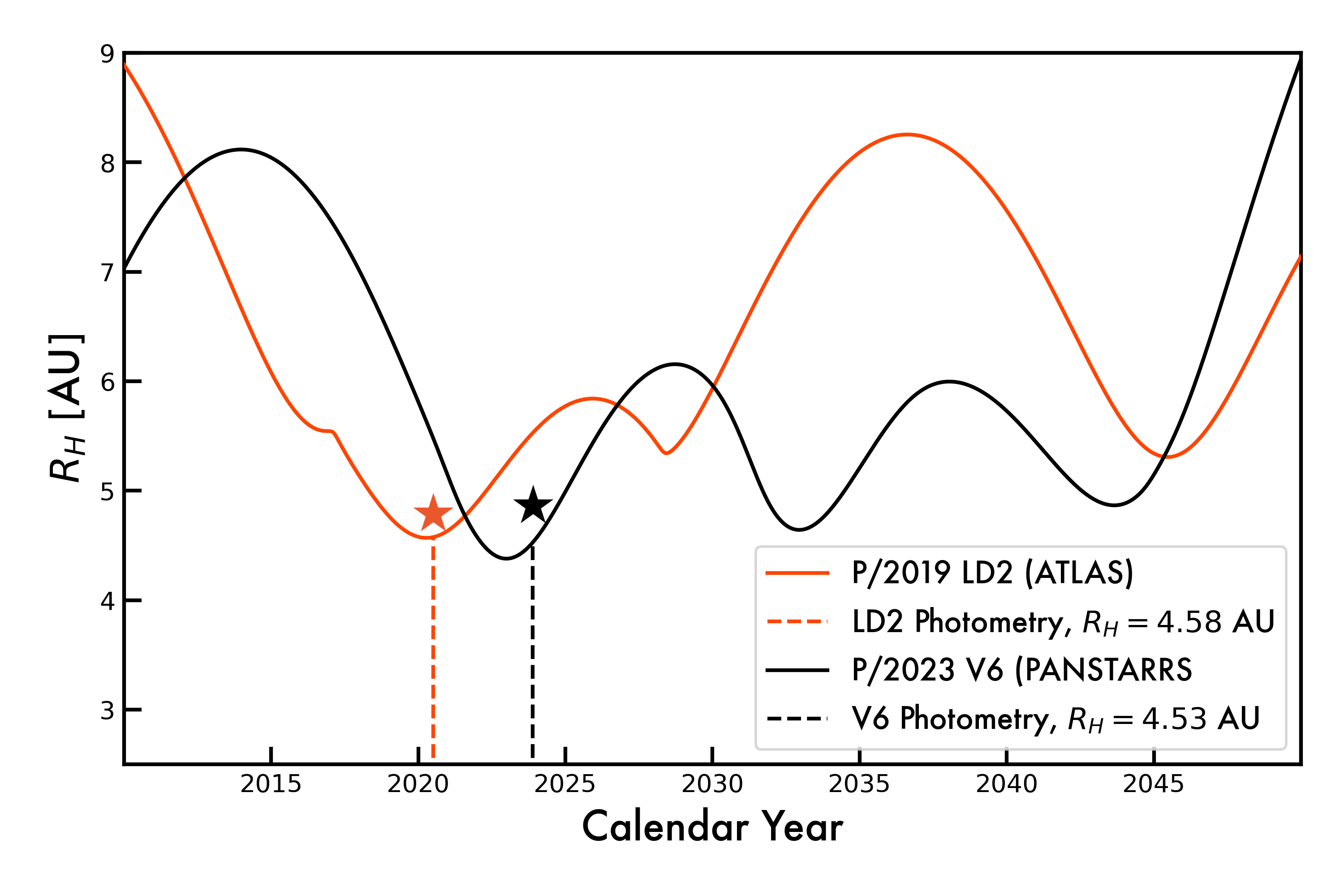}
\caption{The evolution of heliocentric distances $R_H$ for P/2023 V6 (PANSTARRS, black) and P/2019 LD2 (ATLAS, orange) are shown. Two dashed vertical lines are added in the same colors to indicate at what heliocentric distances their $Af\rho$ values were measured with same-colored stars placed above the intersection with the $R_H$ curves. The two objects were both observed shortly after their first perihelion passage after becoming co-orbital, and they both were previously in colder orbits exterior to Jupiter. The similarities begin to end one orbital period after their close encounters, as LD2 is scattered outwards (the sudden change in direction it undergoes in 2028) while V6 is moved into a lower eccentricity orbit which then is scattered outwards one orbital period after that.
}
\label{fig:r_plot}
\end{figure}

P/2019 LD2 (ATLAS) is the only other known active temporary Jupiter co-orbital, and is the most natural comparison for V6. First discussed at the end of Section \ref{sec:obs}, LD2 is radically more active and brighter than V6 is when measured at the same orbital phase of their very similar orbits. We specifically compare their recent heliocentric distance evolutions within the context of when they were observed in Figure \ref{fig:r_plot}. Their long-term orbital evolution is different, but within several years prior to each of the observations the objects received more-or-less the same amount of radiation from the Sun, so their current thermal states should be roughly similar \textit{if} their interior structures and rotational obliquities are roughly similar. Even discounting the reasons that our own $Af\rho$ values might be slight over-estimates, LD2 appears $\sim15\times$ more active using this metric. While neither object displays a large enough coma for morphological analyses, LD2's large-and-persistent tail is clearly at least somewhat different than the roughly-symmetric small circular comae that V6 displays, though the small phase angle differences ($\sim1.2^{\circ}$ for V6 compared to $\sim8.2^{\circ}$ for LD2) might explain some differences in their tail prominences. Both objects show evidence for ongoing, as opposed to sporadic, activity as evidenced by the power law slopes of the brightnesses of their inner comae. We thus explore two effects which might each explain part or all of the differences between these two objects: size and orbital history.

LD2 has a nuclear size of $r_{nuc}<1.2$ km \citep{2021PSJ.....2...48K} assuming a $5\%$ albedo typical for the active Centaurs \citep{2013ApJ...773...22B}, which places it as among the smallest Centaurs known and compatible with a typical size for the JFCs of $r_{avg.}\sim0.65$ km \citep{2017AJ....154...53B}. If V6's ongoing activity is similar to LD2's in term of outflow speed and typical particle sizes, their differences in total mass loss rates should be driven by differences in the amount of total surface areas on their nuclei that are sublimating. If one then assumes that their active fractions -- the fraction of their nucleus that has ongoing mass loss -- are the same, one can then estimate the nuclear size of V6 by scaling its surface area down by the same $\sim15\times$, resulting in a likely nuclear radius of a few hundred meters or less. If LD2's nuclear radius was the same as the upper limit of \cite{2021PSJ.....2...48K}, we would then infer $r\sim0.31$ km for V6 -- about half as wide as a typical JFC. The fact LD2's size is only known to within a reasonable upper limit suggests that the real value for V6's radius could be significantly smaller still. We thus conclude that if the two objects are displaying similar active fractions, V6 has to be very small. Considering that LD2 is already the smallest Centaur to have been discovered, primarily due to its vigorous and long-lasting activity, this would make V6 the smallest Centaur yet known by far. Testing this hypothesis would require high-SNR and high-resolution imaging of V6's inner coma with the hopes of detecting its nucleus directly, but we remind the readers that Hubble \citep{2021AJ....161..116B} was not able to detect LD2's nucleus through its coma. While V6's coma is almost certainly less thick than LD2's, a smaller nucleus might make the problem just as challenging. Lastly, even if V6 had such a small nucleus, the fact that it was detected during a period of stable activity means that larger objects at the same distance should be proportionately easier to detect. In other words, if detecting V6 was possible, where are the other brighter objects in similar orbits? \citet{2020ApJ...904L..20S} argued that objects should be passing through the orbital Gateway at a rate of one per $\sim3$ years if LD2's radius is $\sim1$ km, so there is reason to think there should be more objects yet to be found. In the absence of more evidence that might help discriminate the relative sizes of these two objects, we argue that size alone is unlikely to be the primary driver of V6's weak activity unless the object is smaller than even a typical JFC. 

Adding onto Table \ref{tab:vals}, we compile a list of $Af\rho$ values for comets at similar heliocentric distances ($4.4 < R_H/AU < 4.7$) to V6 and LD2 to assess how their activity compares with those reported in the literature \citep{1999A&A...349..649L, 2001A&A...365..204L, 2003A&A...397..329L, 2005MNRAS.358..641L, 2007MNRAS.381..713M, 2021PSJ.....2...48K}. We note explicitly that these different observers used different techniques and filters to obtain their values, so small differences between observers are not necessarily very meaningful. Even within this narrow range of heliocentric distances exterior to the water ice line, the $Af\rho$ values vary by about a factor of a hundred. LD2 is clearly the most active of the group, with the two runners-up being 74P/Smirnova-Chernykh \citep{1999A&A...349..649L} and the two primary fragments of P/2004 V5 (LINEAR-Hill) \citep{2007MNRAS.381..713M}. 74P is a low-eccentricity comet in the outer Main Belt ($q=3.55$ AU, $Q=4.78$ AU) that is classified as a Quasi-Hilda Comet (QHC, see \citealt{2006A&A...448.1191T}). QHCs are a population of comets temporarily captured into the 4:3 interior resonance with Jupiter, which is a population thought to be dynamically linked to the Jupiter co-orbitals and low-perihelion Centaurs as all of the populations undergo repeated close encounters with Jupiter. Objects can move between these ``near Jupiter" populations, but moving inwards past Jupiter is generally easier than moving outwards, so the QHCs would be assumed to be on average more thermally evolved than the innermost Centaurs. 74P that does not appear to have been `further in' recently, even if QHC orbits are not stable on gigayear timescales, and has a nuclear radius of $r_{nuc}=2.23\pm0.04$ km, or at least double that of LD2. If one assumes that the variation in values in Table \ref{tab:vals} is driven by a combination of sizes and histories, then we could infer that 74P is more evolved than LD2 but less evolved than the other JFCs on the list, which is about what one might infer from its current orbit. The two fragments of P/2004 V5 (LINEAR-Hill), A and B, were observed shortly after their parent body fragmented in two -- but it had been discovered as an apparently inactive body in late 2003 as 2003 YM159 \citep{2007MNRAS.381..713M}. While the two fragments were clearly very active as might be expected for objects whose new exteriors were previously buried and insulated inside a larger body, the fragmentation of their apparently inactive parent as it approached perihelion ($q=4.41$ AU, $Q=11.49$ AU) makes any apples-to-apples assessment of how to compare their activity to those of comets in stable sublimation rather challenging.

While not a direct comparison of $Af\rho$ values but instead of one of the parameters that can be derived from it with assumptions (about dust grain sizes and albedos but also their speeds), we can also compare the bulk dust mass loss rates among objects at these heliocentric distances. For LD2, \cite{2021PSJ.....2...48K} estimated $\sim10-20$~kg/s for a range of assumed grain densities $0.5-1.0$~g/cc. The fragments of Comet Shoemaker-Levy 9 (SL9), spanning the likely sizes for LD2, had their estimated mass loss rates range from $6-22$~kg/s at a slightly higher heliocentric distance of $R_H=5.40$ AU \citep{2000Icar..146..501H} based on an assumed $1$~g/cc density. We note that the mass loss rates measured for the SL9 fragments came from a direct numerical model of the coma as opposed to being derived from $Af\rho$, but the similarity in these values is another indication of just how active LD2 is. We return to this later in the Discussion.

\startlongtable
\begin{deluxetable*}{c|c|c|c}
\tablecaption{Distant Comet ($4.4 < R_H/AU < 4.7$) $Af\rho$ Values.}
    \label{tab:vals}
    \centering
    \startdata
        Comet & $R_H$ [AU] & $Af\rho$ [cm] & Reference\\
        \hline
         P/2019 LD2 (ATLAS) & 4.58 & $484\pm20$ & Kareta et al., 2021\\
         74P/Smirnova-Chernykh & 4.61 & $229\pm11$ & Lowry et al., 1999\\
         P/2004 V5 (LINEAR-Hill) A & 4.43 & $220\pm10$ & Mazotta-Epifani et al., 2007\\
         P/2004 V5 (LINEAR-Hill) B & 4.43 & $143\pm7$ & Mazotta-Epifani et al., 2007\\
         65P/Gunn & 4.43 & $133\pm5$ & Lowry and Fitzsimmons., 2001 \\
         P/1993 K2 (Helin-Lawrence) & 4.70 & $56\pm4$ & Lowry et al., 1999\\
         103P/Hartley 2 & 4.57 & $38\pm3$ & Lowry et al., 2003\\
         P/2023 V6 (PANSTARRS) & 4.53 & $32\pm1$ & This work.\\
         P/2023 V6 (PANSTARRS) & 4.56 & $25\pm1$ & This work.\\
         P/2023 V6 (PANSTARRS) & 4.57 & $24\pm1$ & This work.\\
         43P/Wolf-Harrington & 4.43 & $9\pm1$ & Lowry and Fitzsimmons, 2005\\
         129P/Shoemaker-Levy 3 & 4.59 & $4.6\pm0.6$ & Lowry and Fitzsimmons, 2005
    \enddata
\end{deluxetable*}

V6's $Af\rho$ values are among the lowest on the list, being somewhat lower than 103P/Hartley 2 \citep{1999A&A...349..649L} outbound at the same heliocentric distance. Only two comets, 43P/Wolf-Harrington and 129P/Shoemaker-Levy 3, are lower \citep{2005MNRAS.358..641L}. 43P/Wolf-Harrington is a Jupiter Family Comet which notably has time-variable non-gravitational terms (see, e.g., \citealt{2001A&A...368..676K}) driven by either a nucleus with significant surface heterogeneity (such that the volatile-rich surface areas only receive sunlight along portions of the object's orbit) or a significant offset between the time of perihelion and when non-gravitational effects are the strongest (e.g. the assumed time of peak outgassing). Since the object's (first) discovery in the 1920s, its perihelion has dropped from $q=2.43$ AU to $1.36$ AU. 129P/Shoemaker-Levy 3 is a QHC \citep{2006A&A...448.1191T} like 74P mentioned above with an aphelion $Q$ of $4.75$ AU, and it has not been well characterized physically or chemically. Notably, JPL also utilizes a nonstandard non-gravitational solution in their orbit fit for 129P, but in this case a CO-dominated one. While 129P's orbital arc is not as long as 43P's, it has had the opposite trend in the evolution of its perihelion. Discovered with $q=2.81$ AU, a close encounter with Jupiter in 2009 pushed $q$ outwards to $3.91$ AU. We note explicitly that both of these weakly active comets have $Af\rho$ values from \cite{2005MNRAS.358..641L}, who chose their aperture sizes based on when the comet's brightness became indistinguishable from the background as opposed to choosing based on the slope of the spatial profile. This resulted in a slightly larger utilized aperture than we used ($\sim3.6\arcsec$ vs. our $2.16\arcsec$), which would naturally make our values slightly higher even for the same comet \citep{2005MNRAS.358..641L}. To assess this affect, we recalculated the $Af\rho$ value for V6 based on our November 20 observations using a larger $3.60\arcsec$ radius aperture and derived $Af\rho(3.60\arcsec)=23\pm1$ cm, or about $\sim30\%$ smaller. Similar reductions were found for the later two dates (Dec. 20 and 29), such that our lowest values for V6 are only slightly higher than those measured for 43P. In the absence of more knowledge of the properties of these two weakly-active comets, we thus conclude that the only objects in the literature which show clear activity at similar distances but with possibly weaker strength than we have found at V6 are evolved objects which have spent significant amounts of time in the inner Solar System. V6 is thus less active than even some JFCs, but there are a few objects which likely have slightly weaker activity at the same distance.

\subsection{Modern Similarities, Different Histories}
While V6's orbital history might not be deterministic a century or two into the past, limiting our ability to compare its long-term thermal evolution to that of LD2's, this does not mean this more distant past does not play a role in its modern activity. If V6 had spent significant amounts of time in the Inner Solar System (likely two hundred years ago or earlier based on Figure \ref{fig:longterm}) or if it had spent a very long time in an orbit with $q\approx5$ AU, it would be expected that V6 would have less volatiles in its near-surface than if it had only ever been as close to the Sun as it is now.

Even a glance at Table \ref{tab:vals} shows that LD2 is extremely active for its distance with a higher $Af\rho$ value than any other Jupiter co-orbital comet or JFC measured at the same distance. V6 is weakly active and is nearly the least active object in the whole table, with some of the difference between V6 and the less-active objects being driven by differences in methodology. The different sizes, current orbits, and past histories of all these different objects means that a significant spread in activity strength at the same distance is to be expected. For LD2 and V6, the differences in current orbits can be essentially ruled out, and the previous sub-section was used to argue that the differences in their sizes are unlikely to be large enough to explain their different activity levels, leaving only their orbital history as a significant factor. We thus conclude that the difference in activity strengths between V6 and LD2 is most likely to be driven by different thermal histories for these objects. As V6's orbit is further refined, a future dynamical study could test this hypothesis and investigate what kind of orbit the object used to have -- or at least constrain whether or not this happened any time recently. Furthermore, monitoring of V6's activity could  be very useful both to constraining models of its interior and to making a more substantive comparison to LD2. Following V6 outbound as its activity decreases in strength would eventually allow a nuclear size estimate, and thus a more apples-to-apples comparison of $Af\rho$ values with LD2 and these other objects. Beyond a constraint on V6's size (or an actual measurement of LD2's size!), a constraint on their relative volatile productions -- probably only possible with JWST or ALMA -- would be another way to assess their relative thermal states. If V6 was significantly depleted in more volatile ices, indicated by a lower $Q(CO)/Q(CO_2)$ ratio, compared to LD2, this would also support a more evolved state for the object \citep{2023PSJ.....4..208P}.

The comparisons brought up in this paper highlight a fundamental challenge in interpreting the activity of comets in general. While the JFCs and Centaurs are often treated like separate populations with distinct properties, objects move between these groups frequently -- LD2 will do so in less than 40 years \citep{2020RNAAS...4...74K, 2020ApJ...904L..20S, 2021Icar..35414019H} -- meaning that objects in any single orbital class have a variety of physical and dynamical ages. While the nature of the difference in activity strengths at V6 and LD2 will become clearer with more study of both objects, the magnitude of their differences is a clear reminder that comets are a deeply heterogeneous population due to their deeply heterogeneous orbital evolutions. While it is expected that objects `further out' are less evolved than those further in, this is likely only true on average. V6 and LD2 are an excellent example showing that even when all other factors are considered, a factor of up to $\sim15$ or more difference in bulk mass loss rates is to be expected.

All of these arguments rely upon the assumption that differences in activity strength for two objects in identical orbits with identical sizes should be driven by differences in their past orbits -- whichever spent more time inside the water ice line likely has less water ice left, and so on -- but it is worth stating that differences in their bulk composition might also drive some differences in activity. A comet with a high CO-to-H$_2$O ratio like the Long Period Comet C/2016 R2 (PANSTARRS) (see, e.g., \citealt{2019AJ....158..128M}) or the second Interstellar Object 2I/Borisov \citep{2020NatAs...4..867B, 2020NatAs...4..861C}, would not be expected to lose the same volatiles in the same way as a comet of `typical' composition in those same orbits. We do not think this is likely to be a dominant factor in the differences between these two objects by any means, but it does motivate future searches for volatiles towards both of them to see if their comae appear chemically similar.

\subsection{Implications for Galilean Cratering Record}
The implications of another active temporary Jupiter co-orbital discovered may help to explain the perceived lack of small craters on the Galilean satellites \citep{Zahnle1998cratering,Bierhaus2009europa}. The current prevailing theory is that JFC's are responsible for 90\% of the impacting population on the Galileans \citep{Zahnle1998cratering}. If temporary Jupiter co-orbitals and other transient near-Jupiter orbits are effectively populated via JFCs evolving outwards, and not just Centaurs evolving inward, as suggested by both V6's discovery and dynamical work by \cite{2019ApJ...883L..25S} and others, then the size distribution of impactors to the Jovian system must be skewed by the contribution of a dynamically \textit{and} thermally evolved population. Many of these Jupiter-crossing and temporarily co-orbital objects will have already experienced extended periods of time as JFCs where they will receive far more intense solar radiation and thus experience more severe outgassing and fragmentation events.

While JFCs are not as likely to suffer catastrophic breakup events as Oort Cloud comets, they are still susceptible to major outbursts and fragmentation effects (e.g., 17P/Holmes), which in turn decreases the slope of the JFC size frequency distribution at small sizes. The fragments remaining after breakup events are more susceptible to non-gravitational force spin-up due to outgassing, which can disrupt them further \citep{2021PSJ.....2...14S}.  Given the increased likelihood of deep Jovian system encounters for JFCs evolving outward to a temporary Jupiter co-orbital orbit due to low velocity encounters at aphelion, this suggests that impacts to the Jovian system from a thermally evolved JFC population would make smaller craters relative to similarly sized inbound Centaur counterparts \citep{2003Icar..163..102H,2007Icar..191S.586H}; but this population would also be depleted in small objects for the reasons outlined above, effectively further shifting the expected crater size frequency distribution to smaller sizes and truncating it at the smallest sizes, creating a shallower slope of frequency vs. size in log-log space. 

While the origin of cometary activity at these co-orbital distances is interesting in its own right, the detection and characterization of V6 and LD2 as likely having very different thermal histories provides direct evidence for thermally evolved populations passing in and out of JFC orbits through near-Jupiter orbits. This raises the possibility that there are two distinct impactor populations for the Galilean satellites. A numerical investigation into the expected contribution of the thermally evolved JFCs relative to other impactors is beyond the scope of this paper, but is an excellent area for future efforts.
In essence, we argue that if V6's low activity is driven by the object being substantially thermally evolved then there ought to be a significant number of other bodies at similar heliocentric distances that are also evolved. Considering that smaller comets might not survive such thermal evolution -- they erode away instead of going dormant -- this might be a way to decrease the number of small impactors at the Jupiter system, and thus help explain its cratering record.

\section{Summary}
The inner Solar System is where cometary activity is most frequently studied and most deeply understood, but cometary activity clearly begins when objects are further out and much colder. While we are starting to begin to understand how activity might be initiated on the Centaurs specifically\citep{2024ApJ...960L...8L}, many ongoing questions about how exactly the activity is sustained and what processes power it remain unanswered or unclear. A clearer understanding of how comets and Centaurs are altered at larger heliocentric distances is needed to interpret observations of these same objects closer to the Sun. Which properties are altered during passage inwards, and which ones might be preserved from the time of their formation billions of years ago?

In this manuscript, we have described a combined observations-and-dynamics approach to study the recently discovered temporary Jupiter Co-Orbital Comet P/2023 V6 (PANSTARRS). This object is the second known current temporary Jupiter Co-Orbital after P/2019 LD2 (ATLAS), a very active object \citep{2021PSJ.....2...48K, 2021AJ....161..116B,2021A&A...650A..79L} which will enter the inner Solar system in 2063 after a series of close encounters with Jupiter in the intervening decades \citep{2020RNAAS...4...74K, 2020ApJ...904L..20S, 2021Icar..35414019H}. The primary goals for our study were to understand V6's recent orbital history, its modern activity, and to use the very similar modern orbits of V6 and LD2 as a test case to see how similarly they behave given the same inputs. We found the following:

\begin{itemize}
    \item V6's current activity is very weak compared to that of LD2. Using $Af\rho$ as an activity strength metric and matching methodologies to \citet{2021PSJ.....2...48K}, we find V6's current activity to have $Af\rho = 32\pm1$ cm or less compared to $Af\rho=484\pm20$ cm for LD2 measured at the same orbital phase. We argue that V6 is unlikely to be radically smaller than LD2, and thus V6's active fraction must be smaller than LD2's.
    \item V6, like LD2, is only a transient co-orbital of Jupiter. This phase of V6's orbit persists between late 2020 and 2044, or two Jupiter orbits, which is twice as long as LD2's co-orbital phase (2017-2029). While the recent orbital evolutions of both bodies share many commonalities, their longer-term histories begin to look quite different. For at least a century or so prior to their respective close encounters, V6 has had a slightly lower perihelion than LD2 -- its pre-encounter orbit is likely better classified as a `high perihelion JFC' than LD2's pre-encounter Centaur orbit.
    \item We conclude that in order for V6 to be so much less active than LD2 at the same heliocentric distance in such a similar orbit, it must be comparatively much more devolatilized and physically `older' than LD2. Given two objects in very similar orbits with very similar recent orbital evolutions, there is at least a factor of $\sim15\times$ in activity strengths between objects in stable and ongoing activity. We use this to discuss the heterogeneity of the JFC population at large, compare with the similarly aged 39P/Oterma \citep{2023PSJ.....4..208P}, and to motivate future observations of both objects.
    \item Objects smaller than V6 or LD2 might have eroded away instead of devolatilizing their outer layers and significantly lowering their active fraction as we have proposed for V6. Given that the JFCs approach the Jovian system with slower relative velocities near aphelion, we argue that the physical evolution of the JFCs -- and specifically the fact that smaller ones might disrupt instead of go dormant -- might explain a perceived lack of small craters on the Galilean satellites \citep{Zahnle1998cratering, Bierhaus2009europa}.
    \end{itemize}

\begin{acknowledgments}
TK was supported in part by the Lowell Observatory Marcus Comet Research Fund.
KV acknowledges support from NASA (grants 80NSSC23K1169 and 80NSSC23K0886).
RS is supported by NASA SSO award 80NSSC20K0670.

\end{acknowledgments}

\vspace{5mm}
\facilities{LDT(LMI)}

\software{PhotometryPipeline, Rebound, NumPy \citep{2020Natur.585..357H}}


\bibliographystyle{aasjournal}

\end{document}